\documentclass[12pt,preprint]{aastex}

\shorttitle{Evolution of torus covering factor}
\shortauthors{Minfeng Gu}

\begin{document}


\title{The evolution of dusty torus covering factor in quasars}

\author{Minfeng Gu}
\affil{\rm Key Laboratory for Research in Galaxies and Cosmology, Shanghai Astronomical\\ Observatory,
    Chinese Academy of Sciences, 80 Nandan Road, Shanghai 200030, China}
\email{gumf@shao.ac.cn}

\begin{abstract}
We have assembled a large sample of 5996 quasars at redshift $2.0\le z \le2.4$ (high-z) or $0.7\le z \le1.1$ (low-z) from SDSS data release nine and seven quasar catalogs. The spectral energy distribution (SED) of quasars were constructed by collecting WISE, UKIDSS, and GALEX photometric data in addition to SDSS, from which the IR luminosity at $1-7~ \mu \rm m$ and bolometric luminosity at $\rm 1100~\AA - 1 ~\mu m$ were calculated. A red tail is clearly seen in the distributions of the spectral index in $\rm 1100~\AA - 1 ~\mu m$ both for high-z and low-z sources, which is likely due to red or reddened quasars. The covering factor of dusty torus is estimated as the ratio of the IR luminosity to the bolometric luminosity. We found significant anti-correlations between the covering factor and bolometric luminosity, in both high-z and low-z quasars, however they follow different tracks. At overlapped bolometric luminosity, the covering factor of high-z quasars are systematically larger than those of low-z quasars, implying an evolution of covering factor with redshift. 

\end{abstract}


\section{Introduction}

In the unification scheme of active galactic nulcei (AGNs), dusty torus plays an 
important role in the diversity of observational phenomena \citep{ant93,urr95}. In
Type 1 AGNs, observers can directly see the central nuclei and broad line
region (BLR), which are however obscured by the dusty torus in Type 2 AGNs. Although a lot of efforts have been made, still
little has been known on the geometry, dynamics and evolution of dusty torus \citep{eli08}.

The covering factor of dusty torus can be estimated by the fraction of type 2 AGNs
in certain samples, and a mean value of $\sim0.6$ has been found \citep{law10}. 
In spectral energy distribution (SED) of AGNs, generally two bumps can be clearly
seen \citep{elv94}. The first bump at optical/ultraviolet region is usually believed to
be the thermal emission from accreiton disk, while the second bump at infrared region
is thought to be reprocessed emission from dusty torus by absorbing the centeral
nuclei emission. Therefore, the covering factor of dusty torus can be measured by the
ratio of the torus infrared luminosity to the bolometric luminosity. 

Largely limited by available IR data, the covering factor has been investigated only for 
a few quasar samples with rather limited source number several years ago \cite[e.g., ][]{cao05,mai07}, from which the 
significant anti-correlations have been found between covering factor and bolometric luminosity, and 
black hole mass. The breakthrough comes with all-sky data release from 
$Wide-field~ Infrared ~Survey ~Explorer$ \citep[WISE, ][]{wri10}. The WISE photometric 
data in near- and mid-infrared bands can be used to study the IR emission as well as the dust
covering factor for large AGNs samples \citep{mor11,
cal12,ma13,ros13}. However, previous works are all at redshift of $z<2$, and AGNs at
low and high redshift were usually mixed together \citep{mai07,mor11,ros13}, in which the 
evolution effects cannot be ignored. Although some works investigated covering factor in quasars at narrow redshift ranges 
by using WISE data \citep{cal12,ma13} , there is a lack of comparing covering factor between quasars at
high and low redshift. The recent quasar catalog from SDSS data release nine (DR9) consists of 
large number of quasars at $z\ge2$ \citep{par12}, which enables us to explore not only the covering factor at $z\ge2$, but also the evolution of coverign factor when comparing to $z<2$
quasars by combining with previous SDSS quasar catalogs. Moreover, the dependence of covering factor on 
luminosity can be further studied in narrow redshift ranges by separating the coupling between luminosity and redshift. 
%The evolution of CF by comparing high-z and low-z quasars. 
%The statistically significant results can 
%be obtained using large samples of quasars, although the multi-wavelength cross-correlation 
%will largely reduce the sample size. (by extending the luminosity coverage 
%combining the faint sources in DR9Qs and those bright ones in DR7Qs)

The layout of this paper is as follows: in Section 2, we describe
the source sample; the analysis on the covering factor
are outlined in Section 3; Section 4 includes the discussion; and in
the last section, we draw our conclusions. The cosmological
parameters $H_{\rm 0}=70\rm~ km~ s^{-1}~ Mpc^{-1}$, $\Omega_{\rm
m}=0.3$, and $\Omega_{\Lambda}=0.7$ are used throughout the paper,
and the spectral index $\alpha$ is defined as
$f_{\nu}\propto\nu^{\alpha}$, where $f_{\nu}$ is the flux density
at frequency $\nu$.

\section{The sample construction}

%%%%%%%%%%%% The luminosity difference at high-z 2.2 and low-z 0.9 is about 0.96 dex =2.0*alog10(dl(z=2.2)/dl(z=0.9)) %%%%%%%%%%%%%%%%%%%

\subsection{Multi-wavelength surveys}

With the aim to detect baryon acoustic oscillations (BAO), the five-year program Baryon Oscillation Spectroscopic Survey \cite[BOSS;][]{daw13} of Sloan Digital Sky Survey III
\cite[SDSS-III;][]{eis11}, will obtain spectra of 1.5 million of galaxies and over 150000 quasars at $z >2.15$ in 10000 $\rm deg^2$. The BAO signal will be investigated from the spatial distribution of luminous
red galaxies at $z\sim0.7$, and HI absorption lines in the intergalactic
medium (IGM) as detected in the Lyman$-\alpha$ forest of quasar spectra at $z\sim2.5$ \cite[see][for details]{daw13}.

The SDSS Data Release 9 Quasar (DR9Q) catalog was constructed from the first two years of BOSS operations, which includes 87822 quasars detected over $\rm 3275~ deg^2$, spectrocopically confirmed via visual inspection, having luminosities $M_{i}~ [z = 2] < -20.5$ and either displaying at least one emission line with full width at half maximum (FWHM) larger than $\rm 500~ km ~s^{-1}$ or, if not, having
interesting/complex absorption features \citep{par12}. The robust identification and redshift measurments were performed for each quasar
from the spectra in wavelength region $\rm 3600 - 10500~ \AA$ at a spectral resolution of
$1300 < R < 2500$. The catalog presents the largest sample of quasars
at $z > 2.15$, with a total number of 61931. In addition to five-band $(u, g, r, i, z)$ magnitudes with typical accuracy of 0.03 mag, the catalog contains multi-wavelength data, such as X-ray, ultraviolet, near-infrared, and radio when available \cite[see][for details]{par12}.

%To detect the BAO signal from Lyman$-\alpha$ forest, a surface density of 15 quasars
%with $z \ge 2.15$ per square degree is required, which however needs targeting to %fainter magnitudes than SDSS-I/II. The
%BOSS limiting magnitude for target selection is $r \le 21.85$ or
%$g \le 22$ \citep{ros12}. 
DR9Q catalog includes fainter objects than
SDSS-I/II, since a fainter limiting magnitude in target selection was adopted to obtain high quasar surface density as needed for BOSS to detect BAO signal \citep{ros12}. This has advantage in extending luminosity coverage when combining DR9 quasars with
SDSS-I/II quasars. In this work, we investigate the evolution of covering factor by utilizing large number of quasars at $z\ge2$ in DR9Q catalog, and the 
extended luminosity coverage of low redshift quasars when combining DR9Q with DR7 quasar catalogs.

The SDSS DR7 quasar (DR7Q) catalog consists of 105,783
spectroscopically confirmed quasars with luminosities brighter than
$M_{i}=-22.0$, with at least one emission line having a full width
at half-maximum (FWHM) larger than 1000 $\rm km~ s^{-1}$ and highly
reliable redshifts. The sky coverage of the sample is about 9380
$\rm deg^2$ and the redshifts range from 0.065 to 5.46. The
five-band $(u,~ g,~ r,~ i,~ z)$ magnitudes have typical errors of
about 0.03 mag. The spectra cover the wavelength range from 3800
$\rm \AA$ to 9200 $\rm \AA$ with a resolution of $\simeq2000$ \cite[see][for details]{sch10}.

To study the distribution of covering factor for statistical samples of high-z and low-z quasars, 
we make use of WISE all-sky data release \citep{wri10}, near-IR observations from UKIDSS \citep{law07}, in combination
with optical photometry/spectroscopy from SDSS, and GALEX survey \citep{mar05}.
The WISE all-sky data release consists of imaging of the entire sky in four near- to mid-IR
bands centred on 3.4, 4.6, 12 and 22 $\rm \mu m$ to a depth of 0.04, 0.06, 0.5 and 3.2 mJy ($3\sigma$) 
with an angular resolution of $6''.1$, $6''.4$, $6''.5$, and $12''.0$ in the four bands \citep{wri10}.
The UKIRT Infrared Deep Sky Survey \cite[UKIDSS;][]{law07} is a large-scale near-IR survey with the aim to cover 7500 
square degrees of the Northern sky in four near-IR bands, $Y, J, H$, and $K$ using the UKIRT Wide Field Camera \citep{cas07}. 
The project comprises five surveys the Large Area Survey (LAS), the
Galactic Clusters Survey (GCS), and the Galactic Plane Survey (GPS) to a depth of $K \sim 18$, 
the Deep Extragalactic Survey to $K \sim 21$, and the Ultra Deep Survey to $K \sim 23$.
The $Galaxy~ Evolution ~Explorer$ space mission \cite[GALEX;][]{mar05} has performed
an all-sky imaging survey simultaneously in both the far-UV (FUV) and near-UV (NUV)
bands with effective wavelengths of 1530 and 2310 $\rm \AA$, respectively, sensitive 
to $m_{\rm AB}\sim21$ in the All Sky Imaging Survey (AIS), and to $m_{\rm AB} \sim25$ 
in the Deep Imaging Survey (DIS).

\subsection{Quasars at $2.0\le z \le 2.4$}

We assembled a quasar sample at $z\sim2.2$ by collecting DR9 quasars in redshift range $2.0\le z \le 2.4$. This redshift range 
is selected to cover the peak of redshift distribution of DR9Q ($z\sim 2.2$) \cite[see Fig. 22 in][]{par12}, thus to have a statistically large 
sample. Moreover, the narrow redshift coverage enables us to separate luminosity-redshift coupling. 
For our chosen redshift range, four WISE wavebands cover $\sim \rm  1 - 7~ \mu m$ in quasar rest frame. 
Although it does not cover 10 $\rm \mu m$ - the quasar SED bump of
silicate-dust emission \citep{hao07,mor12}, it includes the SED peak at short wavelength $\rm \lesssim 5 \mu m$ \citep{lei10,mor12}.
According to the mean quasar SED in \cite{ric06}, the IR luminosity in $\rm  1 - 7~ \mu m$ is about half of the total IR luminosity in $\rm  1 - 100~ \mu m$  ($L_{\rm 1-100\mu m}=2.03~L_{\rm 1-7\mu m})$. In this work, the IR emission of dusty torus is measured in $\rm  1 - 7~ \mu m$ (see section 3.1).  
%Therefore, the IR luminosity estimated from WISE data are significant in measuring the bulk emisison of dusty torus, albeit perhaps mainly for the hot dust.

The WISE photometry is given in DR9Q catalog by cross-correlating with
the WISE All-Sky Data Release using a matching radius of 2.0 arcsec \citep{par12}. There are 25607 quasars at $2.0\le z \le2.4$ in DR9Q catalog. 
 To estimate the IR luminosity, we require the detections in all four 
WISE wavebands, resulting in 7971 quasars. The further matching with all five surveys in UKIDSS data release eight (DR8) 
in a search radius of 2 arcsec, yielding a sample of 2056 quasars with detections in all four bands 
of UKIDSS ($Y, J, H$, and $K$), which is called high-z quasars hereafter.  

\subsection{Quasars at $0.7\le z \le 1.1$}

In addition to the peak at $z\sim2.2$, there are two peaks at $z\sim0.8$ and $z\sim1.6$ 
in the redshift distribution of DR9Q catalog, which are due to known degeneracies in the 
SDSS color space \citep{par12}. To explore the quasar properties at distinctive redshift from 
high-z quasars, the low redshift range was selected as $0.7\le z \le 1.1$. This redshift region
covers the peaks at $z\sim0.8$ for DR9Qs, and there are also numerous quasars from DR7Q catalog 
in this range. Therefore, the quasar properties can be studied in a wider luminosity range by 
combining DR7Qs and fainter objects in DR9Q catalog.
 
The quasars at $0.7\le z \le1.1$ were firstly selected from DR9Q 
catalog. There are 10392 quasars at $0.7\le z \le1.1$ in 
DR9Q catalog, of which 6759 quasars have detections in all four WISE bands within
2 arcsec match radius to all-sky data release. The further requirements of detections in all UKIDSS bands
within 2 arcsec searching radius reduce the source number to 1574. To have similar rest frame
wavelength coverage in ultraviolet as high-z quasars, we request GALEX NUV and FUV photometry
by cross-correlating SDSS with GALEX general release 6 and 7 (GR6/GR7) within 3 arcsec.
This radius is recommended for matches
between GALEX and ground-based optical catalogs \citep{bud08}, and it has been used in various works \cite[e.g.][]{che09,bia11,gez13}. 
The matches with GALEX give 851 quasars with both NUV and FUV detections.
%\cite{bia11} argued that for a larger match radius (e.g. 4.2 arcsec), the fraction of multiple matches increases significantly, 
%relative to the total number of matches, and so does the incidence of spurious matches. 
These quasars were complemented with quasars at $0.7\le z \le1.1$ in DR7Q
catalog \citep{sch10}. There are 16695 quasars at $0.7\le z \le1.1$ in DR7Q catalog. 
Matching with UKIDSS DR8 within 2 arcsec gives 4137 quasars, and further searching 
WISE within 2 arcsec yields 4093 sources. The request of GALEX data within 3 arcsec
finalizes a sample of 3125 quasars. After excluding 36 quasars already identified in DR9Q catalog, DR7 low redshift 
sample contains 3089 quasars. The combined low-z sample thus consists of 3940 quasars in total.

\section{Results}

\subsection{The IR and bolometric luminosity}

The SED of each quasar was constructed with available multi-wavelength
datapoints as mentioned in previous section, i.e. data from WISE, UKIDSS, SDSS, and GALEX surveys.
The near-IR UKIDSS, SDSS, and GALEX data were firstly corrected for Galactic extinction using the 
reddening map of \cite{sch98} and the extinction law of \cite{car89}. These data, together with WISE data, 
were directly coverted to the rest luminosity at the rest frequency. The SED examples are shown in Fig. 
\ref{sed}, in which it can be seen that GALEX data were added in low-z quasars to have same 
wavelength coverage as high-z sources at UV region.

The IR emission is thought to result from optical-UV photons reprocessed by surrounding torus dust.
For our samples, IR luminosity is integrated in the rest frame wavelength region 
covered by WISE and UKIDSS. To have a uniform wavelength coverage, the 
long-wavelength end is set to 7 $\rm \mu m$, which approximately corresponds to the 
observed WISE band $22~ \rm \mu m$ at highest redshift $z=2.4$. The low-wavelength limit is set at $1~ \rm \mu m$, 
which separates accretion disk emission from torus emission in composite AGN SED \cite[e.g.][]{elv94}. 
The luminosity at $7~\rm \mu m$ was estimated by power-law interpolation or extrapolation with WISE 12 and 22 
$\rm \mu m$ data for both high-z and low-z quasars. In contrast, the luminosity at $1~\rm \mu m$ was calculated 
by extrapolation with WISE 3.4 and 4.6 $\rm \mu m$, WISE 3.4 $\rm \mu m$ and UKIDSS $K$ data, for high-z and 
low-z quasars, respectively. The IR luminosity $L_{\rm IR}$ was then integrated between $1~\rm \mu m$ and 
$7~\rm \mu m$ by directly linking the luminosity at $1~\rm \mu m$, $7~\rm \mu m$ and the observed datapoints with 
a power-law in each frequency interval. Although the estimated IR luminosity does not extend much into mid-IR, it 
contains a significant portion of overall torus emission, as indicated by the short-wavelength peak 
in the intrinsic AGN SED of \cite{mor12}.

Similar to IR luminosity, the bolometric luminosity for our sample is integrated in optical-UV wavelength
region covered by all quasars. The blue end of quasar spectra at $z>2.2$ are subject to 
absorption by IGM, since the detection of characteristic scale imprinted 
by BAO at $z\sim2.5$ rely on the observed $\rm Ly-\alpha$ forest 
\citep{par12,daw13}.  Based on the composite spectrum of DR9 quasars, we select $\rm 1100~\AA$ 
as short-wavelength limit, above which quasar spectra are severely contaminated by $\rm Ly-\alpha$ forest \citep{par12}. 
The luminosity at $\rm 1100~\AA$ is interploated or extrapolated from SDSS $u$ and 
$g$ data for high-z quasars, while it is from interpolation with GALEX NUV and FUV data in low-z objects. 
The luminosity in $\rm 1100~\AA$ - 1$\rm \mu m$ was then integrated between $1~\rm \mu m$ and 
$\rm 1100~\AA$ by directly linking the luminosity at $1~\rm \mu m$, $\rm 1100~\AA$ and the observed datapoints with 
a power-law in each frequency interval.
In this work, we call the power-law integrated luminosity in $\rm 1100~\AA$ - 1$\rm \mu m$ as bolometric luminosity, which
can be transfered to real bolometric luminosity in 1 $\rm \mu m$ - 10 keV by multiplying a factor of $\sim$ 1.62 deduced from mean quasar SED of \cite{ric06}.
%The bolometric corrections is $L_{\rm bol}~=~1.62~L_{\rm 0.11-1\mu m}$ and 
%$L_{\rm bol}~=~3.11~\nu L_{\nu, \rm 3000\AA}$ \citep{ric06}.

\subsection{SED shape}

We investigated the shape of optical-UV SED, by fitting the datapoints between $1~\mu \rm m$ and 
$1100\rm ~\AA$ with a power-law, from which the spectral index were obtained. The distributions 
of spectral index are shown in Fig. \ref{spi} for both high-z and low-z quasars.
We found that low-z quasars appear slightly bluer than high-z sources, with median spectral index of
-0.33, and -0.48, respectively. It can also be seen that the spectral index distributions are clearly asymmetric, with a red rail in both populations. 
In this work, we tentatively define a red quasar with $\alpha<-1.0$, corresponding to a declined SED in $\rm 1100~\AA$ - 1$\rm \mu m$. 
We found that 234 high-z and 159 low-z quasars are red quasars (see examples in Fig. \ref{sed}), which are likely to be intrinsic red quasars, or reddened quasars due to extinction \cite[e.g.][]{ric03}. 

Interestingly, the bluest SED has a spectral index of 0.29, which is close to $\alpha=1/3$ prediction of standard thin disk 
\citep{sha73}. We found that four high-z and six low-z quasars have spectral index $\alpha>0.2$. 
All these sources could be likely explained with standard thin disk. 

\subsection{Analysis on the covering factor}

The relationship between the IR luminosity and the bolometric luminosity is shown in Fig. \ref{irbol}. 
It is clearly seen that the luminosity of low-z quasars spans about 1.5 orders of magnitude, with DR9 sources
extending to low luminosity region, as expected from their faintness \citep{par12}, while the luminosity of high-z sources
covers about one order of magnitude. The significant correlations are found between IR 
and bolometric luminosity with Spearman correlation coefficient $r_{\rm hz}=0.732$ and $r_{\rm lz}=0.893$
both at $\gg 99.99\%$ confidence level for high-z and low-z quasars, respectively. However, they follow different dependences.
The linear fit gives 
\begin{equation}
\rm log~ \it L_{\rm IR}\rm = (0.58\pm0.01)~ log \it ~L_{\rm bol} \rm +(19.13\pm0.48)   % 0.11-1\mu m
\end{equation}
for high-z quasars. For only DR9 low-z quasars, the relation becomes
\begin{equation}
\rm log~ \it L_{\rm IR}\rm = (0.79\pm0.02)~ log \it ~L_{\rm bol} \rm +(9.50\pm0.70).  % 0.11-1\mu m
\end{equation}
In contrast, it is
\begin{equation}
\rm log~ \it L_{\rm IR}\rm = (0.81\pm0.01)~ log \it ~L_{\rm bol} \rm +(8.29\pm0.29)  % 0.11-1\mu m
\end{equation}
for all low-z sources. 

We here define covering factor (CF) as the ratio of the IR to bolometric luminosity.
The significant anti-correlations are found between CF and the bolometric luminosity with Spearman correlation 
coefficient $r_{\rm hz}=-0.690$ and $r_{\rm lz}=-0.431$ both at $\gg 99.99\%$ confidence level for high-z and low-z quasars, 
respetively (see Fig. \ref{irbol}). The linear fit yields 
\begin{equation}
\rm log~ \it L_{\rm IR}/L_{\rm bol}\rm = (-0.42\pm0.01)~ log \it ~L_{\rm bol} \rm +(19.29\pm0.48) % 0.11-1\mu m
\end{equation}
for high-z quasars. For only DR9 low-z quasars, the relation becomes
\begin{equation}
\rm log~ \it L_{\rm IR}/L_{\rm bol}\rm = (-0.22\pm0.02)~ log \it ~L_{\rm bol} \rm +(9.98\pm0.70). %0.11-1\mu m
\end{equation}
In contrast, it is
\begin{equation}
\rm log~ \it L_{\rm IR}/L_{\rm bol}\rm = (-0.19\pm0.01)~ log \it ~L_{\rm bol} \rm +(8.53\pm0.29) %0.11-1\mu m
\end{equation}
for all low-z sources, which is marginally in agreement with that of \cite{ma13} in similar redshift range.
Although CF follows different dependences on bolometric luminosity, most of quasars in both populations are similarly in CF range of
$10^{-0.5} - 10^{0.2}$. 
%the CF in high-z quasars are systematically higher
%than those in low-z objects, with a median log CF value of -0.14 and -0.25, respectively. 

We collected black hole mass for DR7 low-z quasars from \cite{she11}. In addition, we  % 3083, 72 low-z and 1932 high-z
calculated black hole mass of DR9 quasars using the same empirical relation as did for DR7 low-z quasars in \cite{she11}, which utilizes the luminosity at 3000 $\rm \AA$ and FWHM of Mg II. For our DR9 quasars, the luminosity at 3000 $\rm \AA$ was obtained from interpolation between two adjacent datapoints, and FWHM of Mg II lines 
was tentatively adopted as the sum of the blue and red half width at half maximum (HWHM) provided in DR9Q catalog \citep{par12}.
In Fig. \ref{cfmbh}, we investigate the relationship between CF and black hole mass, and Eddington ratio. 
The Eddington ratio is estimated as $1.62 L_{\rm bol}/L_{\rm Edd}$, in which $1.62 L_{\rm bol}$ is the real
bolometric luminosity in 1 $\rm \mu m$ - 10 keV transferred from $\rm 1100~\AA$ - 1$\rm \mu m$ bolometric luminosity with a correction deduced from mean quasar SED of \cite{ric06}, 
and $L_{\rm Edd}$ is the Eddington luminosity.
There are significant anti-correlations between CF and black hole mass with Spearman correlation 
coefficient -0.209 and -0.294 at confidence level $\gg 99.99\%$ for low-z and high-z quasars, 
respectively. However, there is no strong correlation when combining two populations. While there is 
strong anti-correaltion between CF and Eddington ratio with Spearman correlation coefficient -0.217 
at confidence level $\gg 99.99\%$ in high-z quasars, there are no strong correlations in both low-z 
quasars and combined quasars. In case of low-z quasars, our results are in agreement with \cite{ma13}, also
consistent with \cite{cao05} for Palomar-Green quasars. 

\subsection{Evolution of covering factor}

In flux limit surveys, quasars at high redshift systematically have higher luminosity than objects at low redshift, 
which can be clearly seen from Fig. \ref{irbol}. Moreover, the significant anti-correlations apparently present between
CF and bolometric luminosity, for both high-z and low-z quasars. To avoid luminosity-redshfit coupling, 
we select an overlapped bolometric luminosity range log $L_{\rm bol}=$ $\rm 45.8 - 46.2~ erg~ s^{-1}$, to study the 
difference of covering factor between high-z and low-z quasars. The distributions of CF are displayed in Fig. \ref{cfhl} for 778 low-z and 
966 high-z quasars in the luminosity range. We found that the covering factor of high-z quasars are systematically 
higher than those of low-z quasars, with median log CF values of -0.32
and -0.06 for low-z and high-z quasars, respectively. The Kolmogorov-Smirnov statistic (KS) test shows a significant difference 
between high-z and low-z quasars in CF distributions, with KS statistic value of 0.696 at 
the probabilities of $\rm P\ll 10^{-4}$ that the considered samples are drawn from the same distribution. The CF
difference between high-z and low-z quasars strongly implies an evolution in CF from high to low redshift.

In Fig. \ref{cfhlmbh}, we show the relationship between CF and black hole mass, and Eddington ratio in the overlapped luminosity range. Apparently, 
high-z sources have higher CF, however they are indistinguished from low-z ones in distributions of black hole mass and 
Eddington ratio, with both populations in the range of $10^8 - 10^{10} ~\rm M_{\odot}$, and 0.01 - 1.0, for black hole mass, and Eddington ratio, respectively. This indicates that the CF difference between high-z and low-z quasars may not be caused by the dependence of CF on black hole mass and/or Eddington ratio. 

\section{Discussions}

\begin{figure}
\includegraphics[scale=0.8]{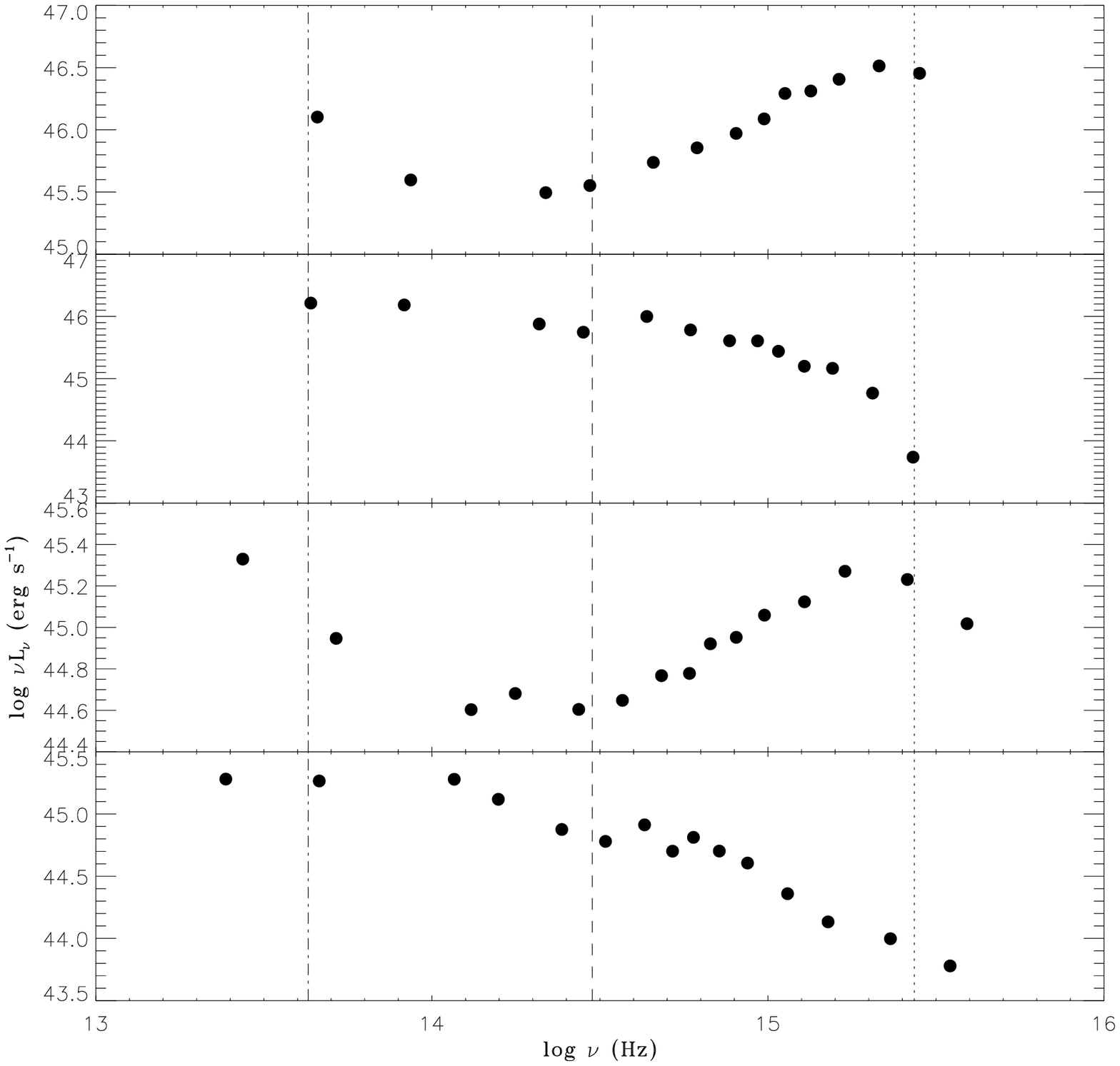}
\caption{Examples of quasar SEDs. From up to bottom: DR9 high-z quasar SDSS 
J000027.01+030715.5 ($z=2.345$); DR9 high-z quasar SDSS J001600.60-003859.2 
($z=2.199$), as a red quasar detected in FIRST; DR7 low-z quasar SDSS J004505.67+002528.1 ($z=1.006$); 
DR7 low-z quasar SDSS J020912.02+004719.0 ($z=0.787$), as a red quasar. In each panel, the dotted, dashed, 
and dot-dashed lines denote the positions of $\rm 1100~\AA$, $1\mu \rm m$ and $7\mu \rm m$, 
respectively. \label{sed}}
\end{figure}

\begin{figure}
\includegraphics[scale=0.8]{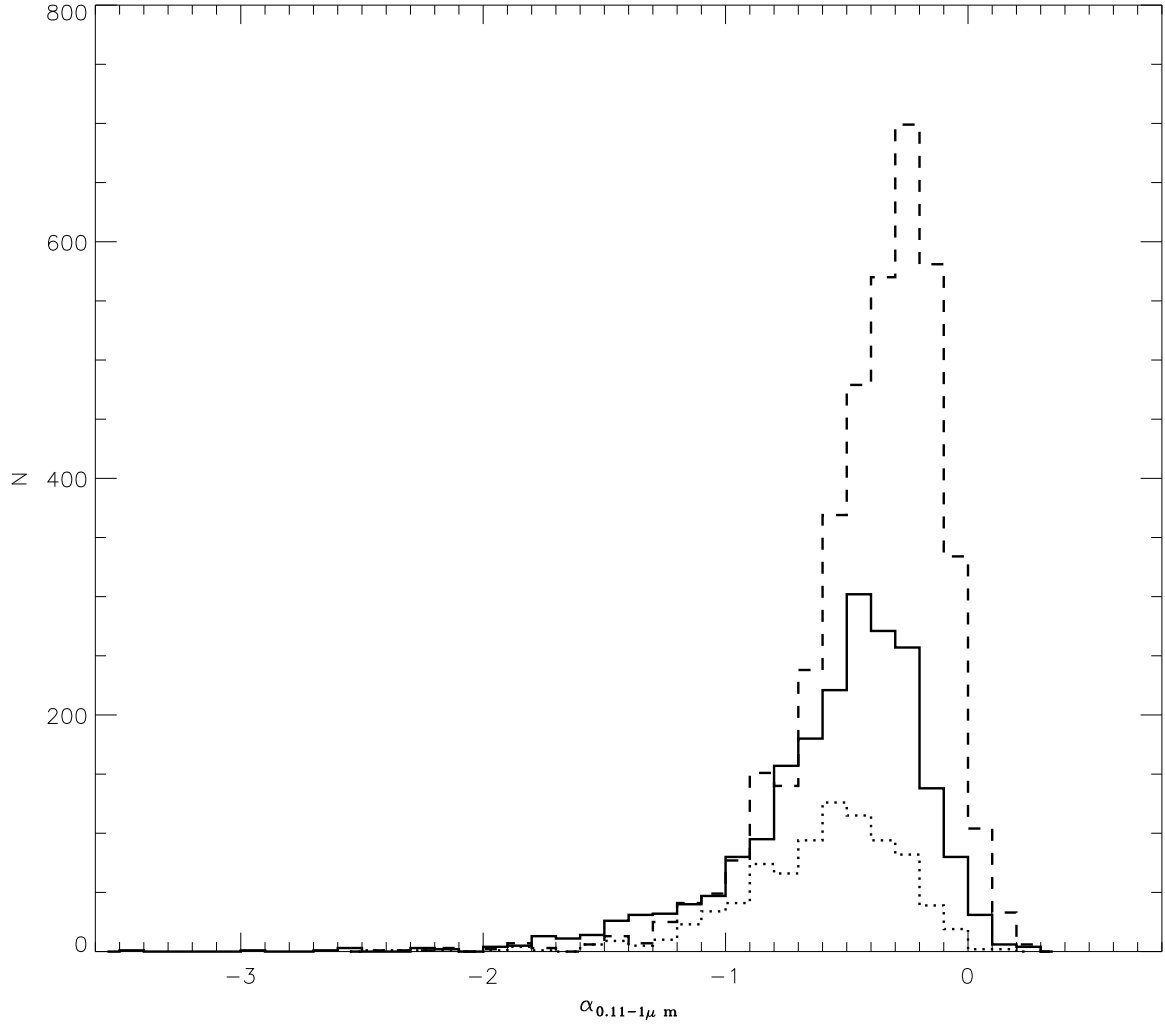}
\caption{The distributions of the spectral index in $0.11-1 \mu \rm m$ region. 
The solid line is for high-z quasars, while the dashed line for low-z sources. The dotted line 
is for DR9 low-z quasars only.
The red tail is clearly seen. \label{spi}}
\end{figure}

As jet emission is usually powerful in radio-loud quasars, more or less it could contribute in both
IR and optical/UV bands. Especially in blazars, the contribution from jet emission is significant, and usually
dominate over the thermal emisson from accretion disk and torus, because their relativistic jets are beamed towards us.
In this case, IR and optical-UV luminosity hardly indicate the torus reprocessed emission, and the accretion disk
emission, thus resulting inappropriate covering factor measurements. We checked radio counterparts of our samples 
in the Faint Images of the Radio Sky at Twenty centimeters (FIRST) 1.4-GHz radio catalog \citep{bec95}.
We found that 79 high-z, and 258 low-z quasars are detected in FIRST, resulting in a total 337 in all 5996 quasars. 
As an example, the SED of FIRST-detected high-z quasar SDSS J001600.60-003859.2 ($z=2.199$) is shown in Fig. \ref{sed}.
The declined SED from IR to UV bands, implies that the synchrotron jet emission may likely be dominated in this source, as in typical blazars \cite[see Fig. 10 in][]{don01}. However, such red SED
has also been found in non FIRST-detected quasars, as shown in Fig. \ref{sed} for DR7 low-z quasar SDSS J020912.02+004719.0 ($z=0.787$). 
This object could be red or reddened quasars, as previously found in SDSS \citep{ric03}. We grouped SEDs of all FIRST-detected quasars together 
by normalizing the luminosity at $1~\rm \mu m$, and found that their group SED is similar to that of non FIRST sources with 
prominent big blue bump. The similar distribution of the spectral index in $\rm 1100~\AA$ - 1~$\rm \mu m$ in two populations 
further support their SED similarities. As stated in \cite{ric06} and \cite{sha11}, the mean SEDs of radio-quiet and radio-loud quasars
in the NUV and IR are quite similar. Therefore, we expect that the FIRST-detected quasars will not affect our statistical results, due to 
both small source fraction ($\sim5.6~\%$), and their similar SEDs as non FIRST objects.
%33 in 851 DR9, and 225 in 3089 DR7 low-z quasars

At different redshift, the observed data at same wavebands actually sample different rest frequencies. The bias caused by
SED sampling may not be severe in optical-UV region, however, there is likely SED undersampling in IR due 
to sparse WISE data. To check the bias caused by IR data sampling, we calculated $1 - 7 ~\rm \mu m$ IR luminosity from the mean quasars SED of \cite{ric06} for $z=2.2$, and $z=0.9$, by using same power-law integration method on WISE data as done for our samples. 
While mean SED gives IR luminosity of $\rm 10^{45.65}~ erg ~s^{-1}$, they are $\rm 10^{45.57} ~erg ~s^{-1}$ and $\rm 10^{45.63} ~erg ~s^{-1}$ 
for high and low redshift, respectively. While the underestimation due to sparse WISE data is not significant, the IR luminosity 
in low-z quasars are basically higher than those of high-z quasars, albeit only slightly. Therefore, if taking IR data sampling into
account, the difference of covering factor can be even larger between high-z and low-z quasars. 
%WISE data actually are at different rest-frame wavelength due to 
%the redshift difference. but, 22 and 12 um of high-z samples at about 6.9 and 3.8 um, 1.4 and 1.1 um, while
%low-z samples at about 11.6 and 6.3 um, 2.4 and 1.8 um, \cite{mor12} IR template shows
%a NIR bump at around 5 um and two bumps at around 10 and 20 um, the separations of the first
%two bumps happens at around 7 um, therefore, the low-z's 7 um could be averagely 
%larger than high-z's, the real difference then would be even larger. the higher CF in high-z
%then is unlikely caused by the sparse datapoints in WISE. also, see QSO IR average spectra in
%\cite{hao07}.

\begin{figure}
\includegraphics[scale=1.0]{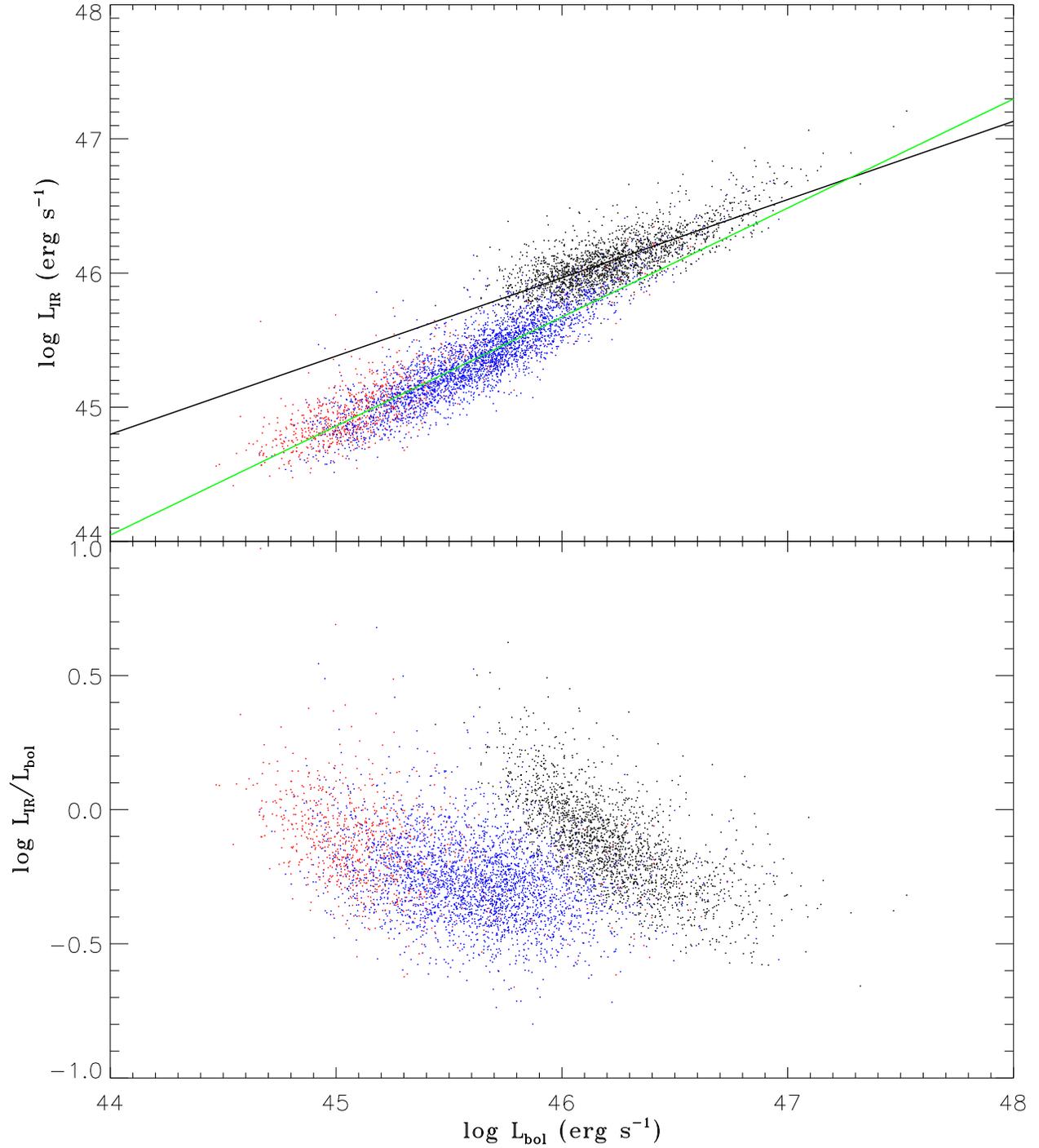}
\caption{Up: the IR luminosity versus the bolometric luminosity; Below: the 
covering factor and the bolometric luminosity. The black dots are high-z DR9 quasars, and 
the red and blue dots are DR9 and DR7 low-z quasars, respectively. The black line
is the linear fit for high-z quasars, and the green one is for all low-z sources. \label{irbol}}
\end{figure}

\begin{figure}
\includegraphics[scale=0.8]{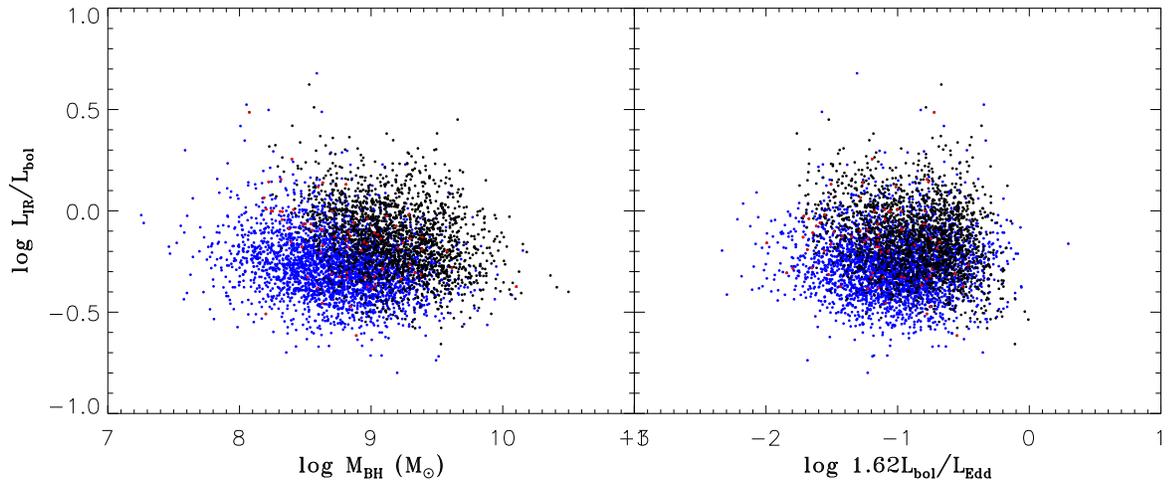}
\caption{$Left$: CF versus the black hole mass; $right$: CF and the Eddington ratio $1.62L_{\rm bol}/L_{\rm Edd}$, 
in which $1.62L_{\rm bol}$ is the bolometric luminosity at 1 $\rm \mu m$ - 10 keV (see text for details). 
The symbols are the same as in Fig. \ref{irbol}. \label{cfmbh}}
\end{figure}

\begin{figure}
\includegraphics[scale=0.8]{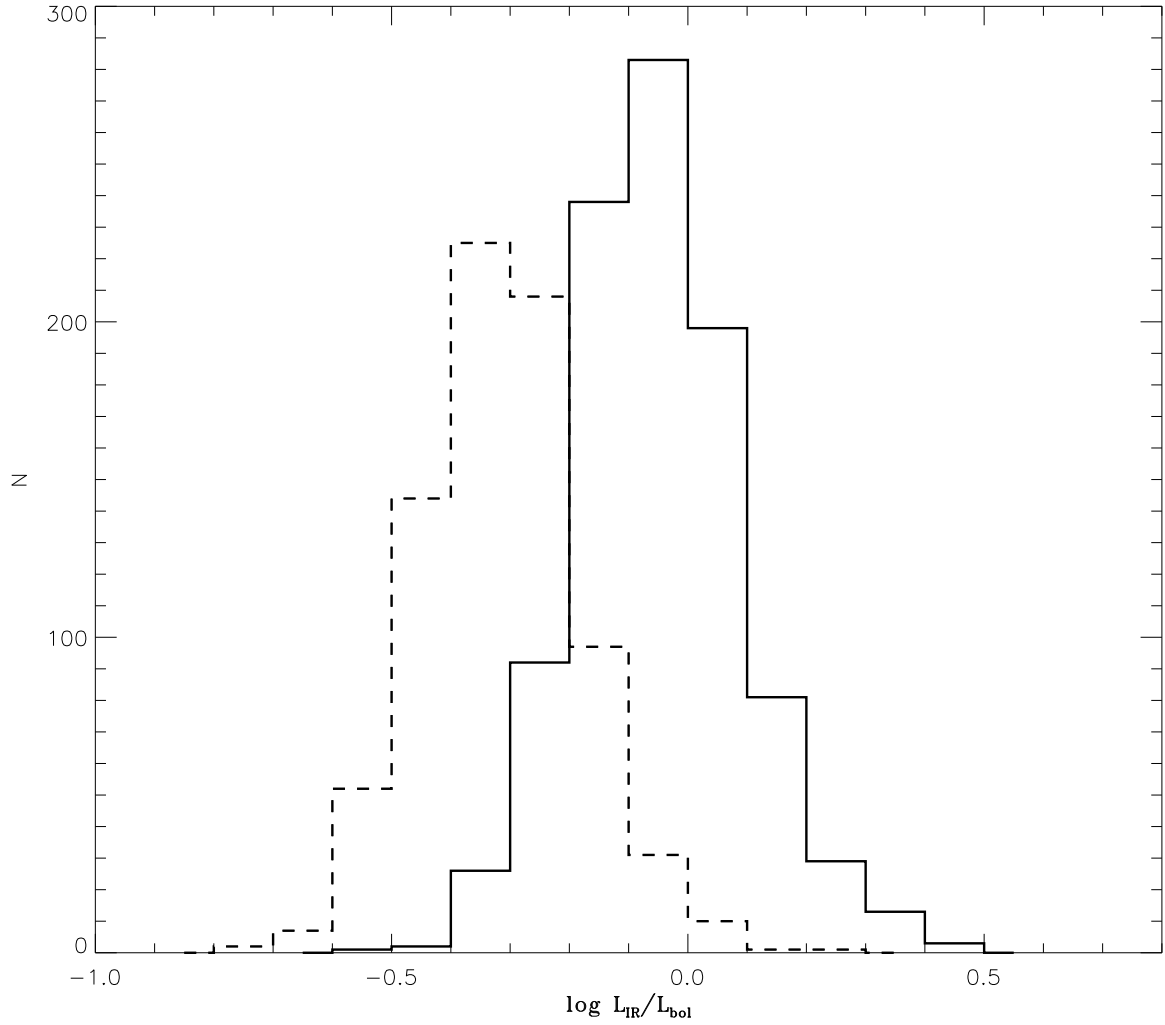}
\caption{The distributions of the covering factor in the bolometric luminosity range $10^{45.8}~-~10^{46.2}\rm ~erg ~s^{-1}$. 
The solid line is for high-z quasars, while the dashed line for low-z sources. \label{cfhl}}
\end{figure}

\begin{figure}
\includegraphics[scale=0.85]{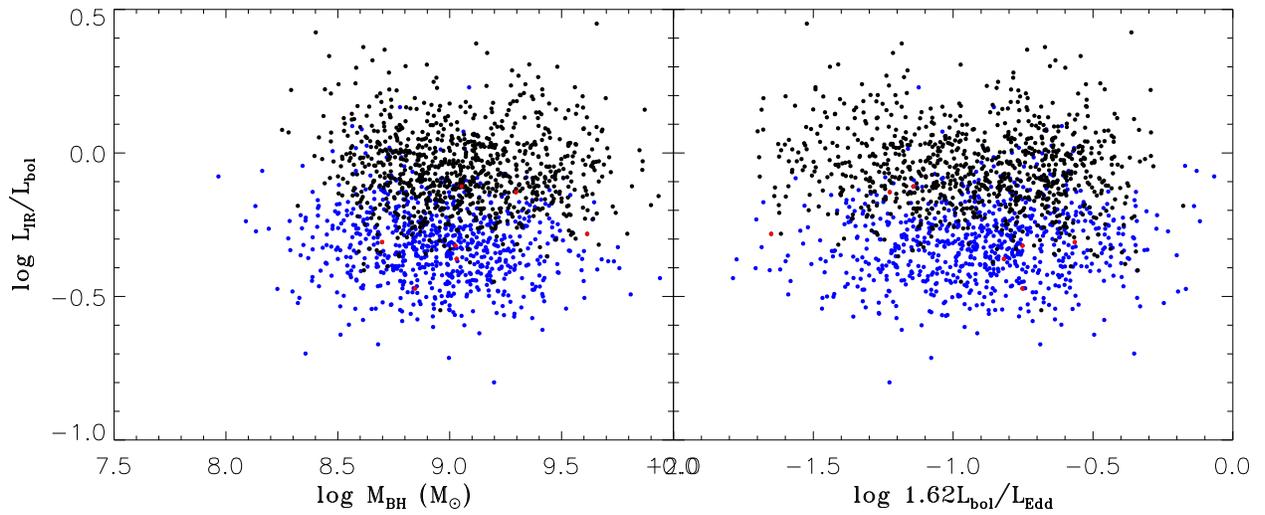}
\caption{For quasars at the bolometric luminosity range $10^{45.8}~-~10^{46.2}\rm ~erg ~s^{-1}$: 
$left$ - CF versus the black hole mass; $right$ - CF and the Eddington ratio, in which $1.62L_{\rm bol}$ is the bolometric luminosity at 1 $\rm \mu m$ - 10 keV. 
The symbols are the same as in Fig. \ref{irbol}. \label{cfhlmbh}}
\end{figure}

In this work, we required detections in all four WISE wavebands to calculate IR luminosity. Therefore, due to shallow WISE sensitivity, especially at 22 $\rm \mu m$, we might have missed a lot of IR faint high-z quasars, although they are detected in SDSS. We found that the distribution of SDSS $r$ magnitude of 
high-z quasars detected in all four WISE bands is significantly different from that of WISE non-detections, with the former systematically brighter than 
the latter. Together with the significant correlation between WISE flux and SDSS $r$ flux in WISE-detected objects, quasars without WISE detections 
are expected to follow the relation of IR and bolometric luminosity of WISE-detected high-z quasars (see Fig. \ref{irbol}). 
%Therefore, there are 
%unlikely lots of high-z quasars with high bolometric luminosity but with low IR luminosity, which is 
%needed to smooth the CF difference between high-z and low-z quasars. 
We revisited the results by only including those
low-z quasars with WISE flux density above WISE flux limit after being moved to $z=2.2$. The CF difference between 
high-z and low-z quasars is still significant at overlapped bolometric luminosity.  
Moreover, we found same result in terms of quasars at overlapped IR luminosity region $\rm 10^{45.6} - 10^{46.0} ~erg ~s^{-1}$. 
All these analysis show that our results are unlikely caused by non-inclusion of IR faint high-z quasars. %the bias in sample selections. 

Quasars usually have strong broad emission lines, which could affect the photometry when emission lines are redshifted into the wavebands. 
In principle, the line contribution to the fluxes at relevant wavebands could be subtracted considering the corresponding spectra 
or the method proposed by \cite{elv12}. In this work, we prefer not to remove
any line contribution, since we mainly focus on the systematic difference between high-z and low-z quasars. However, analysis has been performed
to investigate the influence on our results caused by line contributions. We only considered three strongest emission lines $\rm Ly \alpha$, $\rm H\alpha$, and 
$\rm H\beta$, and found that they are indeed covered by various wavebands in our considered redshift range. While $\rm Ly \alpha$ line
is redshifted into NUV bandwidth for low-z quasars, $\rm H\alpha$ is moved into $J$ for low-z and $K$ for high-z objects, 
and $\rm H\beta$ locates in SDSS $z$ for low-z and $H$ for high-z sources. As stated in \cite{elv12}, the correction of line contribution is dependent on
the line equivalent width (EW) and the bandwidth (see equation (1) in \cite{elv12}). For the typical line EW and bandwidths of related wavebands, we found that the corrections are about 0.1 dex for $\rm H\alpha$ and 0.02 dex for $\rm H\beta$ in both low-z and high-z sources,
and 0.04 dex for $\rm Ly \alpha$ only in low-z quasars. % since $\rm Ly \alpha$ does not influence high-z sources. 
While there are overestimations on the bolometric luminosity in both high-z and low-z quasar due to the inclusion of line contributions from $\rm H\alpha$ and $\rm H\beta$, 
the low-z sources would have additional overestimation of about 0.04 dex from $\rm Ly \alpha$. However, this factor certainly is not enough to
explain the systematical CF difference between high-z and low-z quasars in the overlapped bolometric luminosity. Although the contribution of $\rm H\alpha$
is nontrial, it does not affect our result since we found similar results after excluding the relevant wavebands of $\rm H\alpha$ line. 
Moreover, we checked the results using the bolometric luminosity estimated from the luminosity at $\rm 3000~\AA$, which is interpolated from
two adjacent datapoints and not affected by the line contributions. We found a similar result, i.e. a significant CF difference between high-z 
and low-z quasars. In conclusion, the line contaminations will not affect our results, although we did not perform the corrections.

While the black hole masses of DR7 low-z quasars are collected from \cite{she11}, those of DR9 quasars are obtained by tentatively using 
the FWHM of Mg II lines as the sum of the blue and red HWHM provided in DR9Q catalog \citep{par12}. Thus, the black hole mass
of DR9 quasars are only illustrative, not conclusive. It is still ambiguous whether it is necessary to subtract a narrow line component 
for Mg II, as some works do subtract a narrow Mg II component \cite[e.g., ][]{mcl04} while others do not \cite[e.g., ][]{ves09}.  
The comparision of FWHM of whole Mg II line with FWHM of broad Mg II for DR7 quasars in \cite{she11} shows that the majority of sources ($\sim67\%$) have
difference within 0.02 dex, corresponding to 0.04 dex in black hole mass estimations. Therefore, our black hole mass for DR9 quasars
might be reasonable although not strict. In Fig. \ref{cfmbh}, the black hole masses of DR9 quasars are reasonably in the range from 
$10^8$ to $10^{10}~ \rm M_{\odot}$, with high-z sources having systematically larger values perhaps simply due to the larger luminosity than low-z ones.
%90\% sources have difference of 0.2 dex, resulting a 0.4 dex in Mbh. 80\% within 0.1 dex, and 72\% within 0.05 dex, and 
%67\% within 0.02 dex, and 65\% within 0.01 dex.
Contrary to no correlation in low-z quasars, there is a strong anti-correlation between CF and Eddington ratio in high-z objects.
It is unclear why high-z quasars are different from low-z counterparts. It is worthy noting that the high-z sample size is relatively small, 
and/or the uncertainty in black hole mass estimation is not well quantitatively evaluated. The larger sample from the next version of BOSS 
and future extended BOSS (eBOSS) will help to study these effects. %resolve this issue, in addition to the careful profile analysis of Mg II line.

The quasars in DR7Q catalog were selected as those sources with at least one emission line having FWHM larger than 1000 $\rm km~ s^{-1}$, while DR9 quasars were selected either display at least one emission line with FWHM larger than $\rm 500~ km ~s^{-1}$ or, if not, have
interesting/complex absorption features. Due to different selection criterions, high-z quasars in DR9Q catalog may likely bias towards type 2 quasars, as they may include a fraction of sources with narrower emission lines than typical type 1 quasars. Since type 2 AGNs are thought to have larger covering factor than type 1 AGNs \cite[e.g.][]{eli12}, this selection bias perhaps will at least partly produce the higher CF in high-z quasars compared to low-z quasars, which are mainly from DR7Q catalog. However, we found that almost all quasars at $2<z<2.4$ in DR9Q catalog ($\sim99\%$) have FWHM $\rm >1000 ~km~ s^{-1}$ of at least one of Mg II, and C IV lines. In fact, in our high-z quasar sample, 2048 of 2056 quasars ($\sim99.6\%$) have FWHM $\rm >1000~ km ~s^{-1}$, and there are 962 of total 966 high-z quasars ($\sim99.6\%$) having FWHM $\rm >1000 ~km ~s^{-1}$ in overlapped bolometric luminosity range. When restricting FWHM $\rm >2000~ km~ s^{-1}$ in 960 of 966 quasars, we found a same log CF median value of -0.06, proving the significant CF difference between these broad line high-z quasars and low-z quasars. We further check the results by only including quasars with a narrow range of spectral index in the overlapped bolometric luminosity range. In the range of $-0.3<\alpha<-0.2$, we found that the CF difference remains significant with median log CF values of -0.11, and -0.32 for high-z, and low-z quasars, respectively. The KS test shows a significant difference between high-z and low-z quasars in CF distributions, while there is no strong difference in $\alpha$ distributions. Similar results are found in other $\alpha$ ranges. Finally, we check the results by comparing high-z quasars with only DR9 low-z quasars. The median log CF value is -0.31 for low-z DR9 quasars in overlapped $L_{\rm bol}$ range, which confirms the significant difference compared to high-z quasars. We therefore conclude that the CF difference between high-z and low-z quasars may be less likely caused by the selection bias in DR9Q catalog.

For the first time, the large sample of DR9 quasars enables us to explore the statistical properties of 
the IR luminosity and then the covering factor for quasars at $z>2$. The uniqueness of our sample is that we selected the sources 
with detections from IR all though to UV from various surveys. Although the sample size is much reduced, our sample is large enough for statistical 
investigations. More importantly, the bolometric luminosity can be better constrained with multi-wavelength data in comparison to estimation from UV luminosity 
at single wavelength (usually $\rm 3000~\AA$) or integrated in a narrow UV wavelength range \citep{mor11,cal12,ma13}. 
Due to lack of X-ray measurements, we calculated the bolometric luminosity only in $\rm 1100~\AA$ - 1$\rm \mu m$, which however
represents majority ($\sim 62 \%$) of the overall bolometric luminosity in 1 $\rm \mu m$ - 10 keV according to the mean quasar SED of \cite{ric06}.
Instead of calculating torus IR luminosity using certain models \citep{mor11,ros13}, we obtain the IR luminosity from direct 
power-law integrations, which is basically same to \cite{cal12} and \cite{ma13}. While we have proved that the IR luminosity calculation 
won't bring significant systematic bias when comparing high-z to low-z quasars, it is certainly also true in each populations, 
since the redshift range is rather narrow in each sample.
Benefiting from our source selection, the dispersion of the relationship between IR and bolometric luminosity is much less than that of \cite{ma13} at similar redshift (see Fig. \ref{irbol}), although their IR luminosity is in $\rm 3 - 10~ \mu m$ range.

We have shown that the anti-correlations between the covering factor and 
bolometric luminosity follow different tracks for high-z and low-z quasars, while their covering factor cover similar range (see Fig. \ref{irbol}).
The overlapped bolometric luminosity, actually corrsponds to the high luminosity end of low-z quasars, while it's low luminosity end 
of high-z ones. Therefore, it is quite natural to see a higher covering factor of high-z quasars at overlapped bolometric luminosity.
While the anti-correlation between the covering factor and bolometric luminosity has been found in various occasions and different scenarios have
been proposed \cite[e.g., ][]{mai07,mor11,cal12,ma13}, we found a higher covering factor in high-z quasars, implying an 
evolution of covering factor with redshift. This result is consistent with the findings that the fraction of obscured AGNs increases 
significantly with redshift in hard X-ray selected samples \cite[e.g., ][]{la05,has08}. 
%However, it is unclear why this could happen, especially when their distributions of black hole mass, Eddington ratio are rather
%similar at this luminosity range. In other words, these two populations are basically same in terms of the luminosity, black hole mass, and accretion rate.
According to the merger scenario, the merger triggers the star fromation and quasar activity as well. The quasars finally will blow away 
the surrounding dust and gas, and outshine the host galaxy. At that time, plenty of materials supply to the central accretion process and formation of quasar structures. An increase of the covering factor at high redshift could be naturally expected by the larger gas contents and associated with the 
enhanced star formation rates in high-z galaxies, which was also observed in the host galaxies of high-z AGNs \cite[e.g., ][]{shi09,tac10}.
%Naively, high-z quasars would be relatively young, and then the larger CF of dusty torus may be due to the much more gas and dust around central nuclei. 
%However, it's unclear why this effect is not reflected in the accretion rate, since high-z quasars are expected to have higher accretion rate in this scenario, which is
%contrary to our findings. 
%the origin of torus, lawrence and Elvis 2010
%paper on the warped disk, and also the disk wind model, to explain the results ???

%%%%%% There is unlikely an dependence of covering factor on luminosity, from the work on fractions of type 2 AGNs (Lawrence 2013),
%           This is contrary to our findings.
%           transfered to L_1-100um/L_1um-10kev, the mean values of log CF' are -0.04, and -0.15 for high-z and low-z, respectively.
%           However, according to Hasinger+08 and La Franca+05, the fraction of type 2 AGNs are 0.6 (-0.22 in log), and 0.4 (-0.4 in log) at z~2.2, and z~0.9, respectively.
%           Therefore, there are numbers of quasars in our sample have log CF' >0 (mostly could be red quasars)
%           The log CF' in our samples is larger than Hasinger+08 and La Franca+05, which is likely due to the missing population of compton-thick AGNs even in hard X-ray samples.
%%%%%% see Bianchi et al. 2012AdAst2012E..17BAGN Obscuration and the Unified Model for evolution of covering factor

\section{Summary}

By constructing SEDs for a sample of 5996 quasars at redshift $2.0\le z \le2.4$ or $0.7\le z \le1.1$ from SDSS DR9Q and DR7Q catalogs, the covering factor of the dusty torus is estimated as the ratio of the IR luminosity to the bolometric luminosity. We found significant anti-correlations between the covering factor and bolometric luminosity, both in high-z and low-z quasars, however they follow different tracks. At the overlapped bolometric luminosity, the covering factor of high-z quasars are systematically larger than those of low-z quasars, implying an evolution of covering factor with redshift. 

\acknowledgments
We thank the anonymous referee for constructive comments
that improved the manuscript. We thank Xinwu Cao for the valuable discussions.
This work is supported by the 973 Program (No. 2009CB824800), and by the National Science 
Foundation of China (grant 11073039).
This publication makes use of data products from the Wide-field Infrared Survey Explorer,
which is a joint project of the University of California, Los Angeles, and the Jet Propulsion 
Laboratory/California Institute of Technology, funded by the National Aeronautics and Space
Administration. This study makes use of data from the SDSS (see http://www.sdss.org/collaboration/credits.html).

\clearpage

\end{document}